# Parallel transport modeling of linear divertor simulators with fundamental ion cyclotron heating[*]

A. Kumar[1], J. F. Caneses-Marin[1,2], C. Lau[1] and R. Goulding[1]

[1]Oak Ridge National Laboratory, 1 Bethel Valley Road, Oak Ridge, TN 37831, United States of America

[2]CompX, Del Mar, CA, United States of America

**Abstract**

The Material Plasma Exposure eXperiment (MPEX) is a steady state linear device with the goal to perform plasma material interaction (PMI) studies at future fusion reactor relevant conditions. A prototype of MPEX referred as 'Proto-MPEX' is designed to carry out research and development related to source, heating and transport concepts on the planned full MPEX device. The auxiliary heating schemes in MPEX are based on cyclotron resonance heating with radio frequency (RF) waves. Ion cyclotron heating (ICH) and electron cyclotron heating (ECH) in MPEX are used to independently heat the ions and electrons and provide fusion divertor conditions ranging from sheath-limited to fully detached divertor regimes at a material target. A Hybrid Particle-In-Cell code- PICOS++ is developed and applied to understand the plasma parallel transport during ICH heating in MPEX/Proto-MPEX to the target. With this tool, evolution of the distribution function of MPEX/Proto-MPEX ions is modeled in the presence of (1) Coulomb collisions, (2) volumetric particle sources and (3) quasi-linear RF-based ICH. The code is benchmarked against experimental data from Proto-MPEX and simulation data from B2.5 EIRENE. The experimental observation of "density-drop" near the target in Proto-MPEX and MPEX during ICH heating is demonstrated and explained via physics-based arguments using PICOS++ modeling. In fact, the density drops at the target during ICH in Proto-MPEX/MPEX to conserve the flux and to compensate for the increased flow during ICH. Furthermore, sensitivity scans of various plasma parameters with respect to ICH power are performed for MPEX to investigate its role on plasma transport and particle and energy fluxes at the target. Finally, we discuss the pathway to model ECH in MPEX using the Hybrid PIC formulation herein presented for *kinetic* electrons and fluid ions.

## 1 Introduction

Understanding and controlling Plasma-Material-Interactions (PMI) in future fusion reactors is considered a critical step towards the realization of commercially-viable power-producing fusion reactors. Present-day toroidal fusion confinement devices can explore important plasma physics questions towards this goal but are limited in their pulse duration and ion fluence on Plasma Facing Components (PFCs). To address this limitation, steady state linear divertor simulators have been proposed [1]. In the USA, the response to this need has led to the development of a new linear divertor simulator: the Materials Plasma Exposure eXperiment (MPEX) [2, 3]. The MPEX device is currently under construction at Oak Ridge National Laboratory in the USA. A predecessor experiment, Proto-MPEX, has been used to develop the source, transport, and heating concepts.

[*] This manuscript has been authored by UT-Battelle, LLC, under contract DE-AC05-00OR22725 with the US Department of Energy (DOE). The US government retains and the publisher, by accepting the article for publication, acknowledges that the US government retains a nonexclusive, paid-up, irrevocable, worldwide license to publish or reproduce the published form of this manuscript, or allow others to do so, for US government purposes. DOE will provide public access to these results of federally sponsored research in accordance with the DOE Public Access Plan (http://energy.gov/downloads/doe-public-access-plan).

MPEX will have the capability to expose targets up to $10^6$ seconds and ion fluences ($\sim 10^{31} \, s^{-1}$) not achievable in present-day and planned toroidal fusion devices. MPEX will allow exploration of questions relating to long pulse operation such as testing of novel PFC materials and power exhaust solutions, hydrogenic retention in PFCs under steady-state conditions, etc. A key capability of MPEX will be the independent control of ion and electron temperature via the application of Radio-Frequency (RF) fields at suitably selected cyclotron resonant surfaces. The ion and electron heating R&D needed to develop the linear divertor simulator concept has been carried out in the Proto-MPEX device. References to previous numerical and experimental work on ion and electron heating in Proto-MPEX can be found in references [4-12]. Notable work on the subject of fundamental ion cyclotron heating in open systems include the VASIMR electric thruster [13] and the HITOP device in Japan [14].

MPEX will expose material targets to steady-state plasma conditions that resemble those expected in the ITER divertor. The plasma conditions will be created using Radio-Frequency (RF) power to resonantly heat both electrons and ions via cyclotron interactions. However, these resonant heating schemes will create anisotropic and non-thermal features in the distribution function. Moreover, the open and non-uniform magnetic field used in MPEX will affect the parallel transport of particles and power to the target region due to kinetic and mirror effects during RF heating as demonstrated in reference [11, 12]. The final state of the plasma, subject to all these interactions (RF heating, collisions, kinetic transport, mirror effects, open and non-uniform magnetic field), cannot be fully described with fluid models [15-17] or bounced-averaged kinetic codes [18, 19]. Addressing this transport problem requires solving the Vlasov equation with appropriate particle and energy sources and *coupled* to electromagnetic field equations. A linearized solution to this problem has been carried out to model electron heating in Proto-MPEX subject to 2[nd] harmonic EBW heating [11]; however, the electromagnetic fields and plasma moments were not coupled to the Vlasov equation. Linearization is suitable under certain scenarios but is not applicable in the general case where non-thermal and kinetic effects modify the *bulk* plasma.

The principal goal of the work herein presented is to investigate the parallel plasma transport in Proto-MPEX and MPEX during fundamental Ion Cyclotron Radio-Frequency heating (ICH) with up to 400 kW of power. More specifically, we aim to explore the effect of RF heating on the shape of the ion distribution function at the target, kinetic effects on parallel transport and modification of the background helicon plasma. The transport problem is addressed by solving the Vlasov equation *coupled* to electromagnetic fields [6] using the so-called "hybrid" Particle-In-Cell (PIC) approach [20, 21] and adapted to the present scenario (volumetric particle sources, non-uniform magnetic field, open magnetic field lines, RF heating, coulomb collisions). The term "hybrid", in this context, refers to the process of describing ions kinetically and electrons as a fluid. In the present work, parallel transport is taken to be much greater than radial transport and thus the transport problem is solved along magnetic flux surfaces; as a result, the Vlasov equation is solved by retaining one dimension in physical space (bounce-averaging is *not* used) and two dimensions in velocity space by describing ions using the Guiding-Center (GC) approximation. Moreover, Coulomb collisions are modeled with a Fokker-Planck (FP) operator [11, 22] and RF heating using a quasilinear operator [11].

The structure of this paper is as follows: in section 2, a brief description of Proto-MPEX and MPEX including the magnetic field profiles used is discussed; the hybrid PIC code: PICOS++, is introduced and its framework described in section 3; The various operators (RF heating, Coulomb collisions based on Fokker-Planck equation) used are described in section 4. In section 5, results from PICOS++ modeling to study the plasma parallel transport during ICH heating in Proto-MPEX and MPEX are presented. This includes various validation exercises of PICOS++ simulation results with existing experimental data on Proto-MPEX and predictions of plasma parallel transport for MPEX. The results are discussed in section 6 as well as the relationship to the future operation of MPEX.

## 2  The Material Plasma Exposure eXperiment (MPEX)

MPEX can be divided into the following 5 regions as shown in Fig 1: (1) plasma "Dump", (2) "Helicon" plasma source, (3) Electron Cyclotron heating (ECH), (4) Ion Cyclotron Heating (ICH) and (5) "PMI" (Plasma-Material Interaction) region. Each region has a different requirement for the magnetic field based on its function [3]. This leads to the non-uniform magnetic field presented in Fig.1.

The "Dump" region consists of a large diameter water-cooled target whose function is to terminate and recombine the incident plasma flux during steady-state operation. The magnetic field in this region needs to be diverging in order to spread out and minimize the plasma-induced heat flux. The neutral gas pressure in this region is in the order of 1 Pa.

The "Helicon" plasma source [23, 24] is responsible for producing a low temperature plasma using 13.56 MHz RF fields with up to 200 kW. Based on experimental work in Proto-MPEX, the helicon plasma source operates best at a magnetic field below 0.2 Tesla and magnetic mirrors on either end. The MPEX helicon source is expected to provide around $5 \times 10^{21}$ $s^{-1}$ deuterium ions per seconds to the target when operating at 200 kW and 0.2 T source magnetic field at an ionization fraction close to 0.9 [25].

The ECH region consists of a 70 GHz microwave system with up to 400 kW to drive 2$^{nd}$ harmonic (Electron Bernstein Wave) EBW heating via O-X-B [12]. At these conditions, the resonant magnetic field is about 1.25 T. The ICH system consists of a 4-9 MHz RF system with up to 400 kW to drive beach heating of ions at the fundamental ion cyclotron frequency [2]. In a deuterium plasma and an RF frequency of 8 MHz, the resonant magnetic field is 1.1 T.

The "PMI" region consists of a water-cooled target exposed to the intense plasma flux and a magnetic field of 1 T. Given the high neutral gas pressure in both the "Helicon" and PMI regions (>1 Pa), gas skimmers on either end of the heating regions are used to keep the neutral pressure in the ECH and ICH heating regions at about 0.01 Pa [24] When considering all of the above, the MPEX magnetic field needs to be non-uniform (Fig. 1) to accommodate all the RF heating systems and PMI requirements.

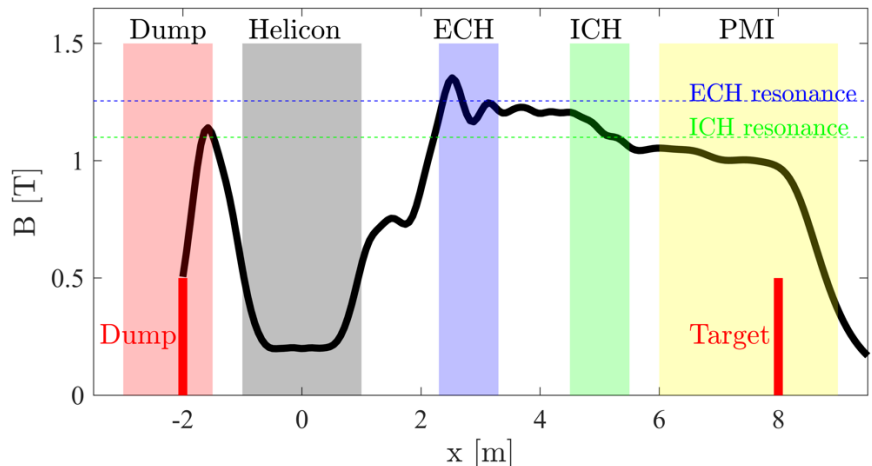

*Figure 1, MPEX is divided into five main regions. Each region has a magnetic field requirement, and this determines the overall magnetic field profile along the device.*

The magnetic field profile and various heating scenarios for Proto-MPEX are discussed in references [11, 26].

## 3  The PIC framework

In this section, we describe the general framework and equation set used in the "hybrid" PIC code which hereafter referred to as "PICOS++" (Particle-In-Cell for Open Systems). PICOS++ is a multi-species, fully-parallel (MPI-openMP), electrostatic, 1D-2V hybrid PIC code used for simulating plasmas in open and non-uniform magnetic field geometries. The public repository for PICOS++ can be found in reference [27].

The Particle-In-Cell (PIC) method is a technique used for self-consistently solving the Vlasov equation together with the electromagnetic fields. The PIC method evolves a collection of super-particles in phase space $(x, v, t)$ according to the equations of motion and subject to the self-consistent electromagnetic fields. The particle phase space distribution is used to calculate charge and current density which become sources to the electromagnetic fields. This cycle is applied repeatedly to advance the phase-space distribution of particles forward in time. The basic PIC cycle is well illustrated in figure 1 of reference [28].

Fully-kinetic PIC codes have been used extensively in the plasma physics community to model processes such as laser-plasma interactions, electric thrusters, basic plasma physics, plasma instabilities, etc. Fully-kinetic PIC codes describe both electrons and ions kinetically. Due to the need to resolve both electron time and length scales, fully-kinetic codes can be computationally intensive especially if physics over ion time scales is required. A common way to reduce computational intensity while retaining ion kinetic physics is to use a "hybrid" PIC approach. The term "hybrid" in this paper refers to the process of describing ions kinetically while electrons as a fluid. More details of the hybrid approach can be found in reference [29-48].

### 3.1 Ion distribution function

The basis of the PIC method is the Vlasov equation (Eq. 1) which describes the time-evolution of the plasma consisting of charged particles which interact via long-range electromagnetic fields. In the form presented in Eq. 1, no collisional processes are present. The effect of Coulomb collisions, RF heating and particles sources are incorporated as source terms on the right-hand side of Eq. 1.

$$\frac{\partial f}{\partial t} + \boldsymbol{v} \cdot \frac{\partial f}{\partial \boldsymbol{x}} + \frac{q}{m}(\boldsymbol{E} + \boldsymbol{v} \times \boldsymbol{B}) \cdot \frac{\partial f}{\partial \boldsymbol{v}} = 0 \qquad \text{Eq. 1}$$

The PIC method solves the Vlasov equation by using a so-called Klimontovich distribution function in Eq. 2, where $N_R$ is the total number of particles in the system modeled. Each particle is described as a delta function in phase-space with a position given by $(\boldsymbol{x}_i(t), \boldsymbol{v}_i(t))$ and evolved in time according to the equations of motion.

$$f_K(\boldsymbol{x}, \boldsymbol{v}, t) = \sum_{i=1}^{N_R} \delta(\boldsymbol{x} - \boldsymbol{x}_i(t)) \delta(\boldsymbol{v} - \boldsymbol{v}_i(t)) \qquad \text{Eq. 2}$$

In all PIC codes, the delta functions are replaced by "shape" functions to reduce shot noise [28]. Different shape functions can be used depending on the application; however, all have the effect of "spreading" the particle's dimension in phase space. All shape functions $S$ must satisfy the property shown in Eq. 3 and effectively behave as a broadened delta function. More details on shape functions can be found in reference [25]. In PICOS++, a triangular-shaped cloud function is used. Details are presented in Appendix 2: Shape and assignment function.

$$\int_{-\infty}^{+\infty} S(x' - x_i) dx' = 1 \qquad \text{Eq. 3}$$

An important assumption used in PICOS++ is that parallel transport dominates over radial transport; hence, particles are restricted to flow along magnetic flux surfaces only. Integrating the distribution function over an annular cross section area $A$ defined by the magnetic flux surface and letting the magnetic field vary along the "x" coordinate leads to the Klimontovich distribution function shown in Eq. 4 where $A(x_i)$ represents the plasma cross sectional area at location $x_i$ and describes the "compression" effect caused by the magnetic field.

$$f_K(x, \boldsymbol{v}, t) = \sum_{i=1}^{N_R} \frac{S(x - x_i(t))}{A(x_i)} S(\boldsymbol{v} - \boldsymbol{v}_i(t)) \qquad \text{Eq. 4}$$

To proceed, the total number of real particles $N_R$ in the plasma system is represented by a much smaller number of so-called computational particles $N_C$. Moreover, each computational particle is given a variable weight $a_i$ to represent an arbitrary number of super-particles. Moreover, using the paraxial approximation for the magnetic field (Eq. 6), the term $A(x_i)$ is calculated using conservation of magnetic flux ($\nabla \cdot B = 0$) which leads to $A(x_i) = A_0 \left(B_0/B(x_i)\right)$, where $A_0$ represents a reference cross sectional area. Hence, the distribution function used in PICOS++ is shown in Eq. 5 where $K = N_R/N_{SP}$ represents the number of real particles $N_R$ for every super-particle, $N_{SP}$ is the total number of super-particles (defined below), $N_C$ the total number of computational particles, $a_i$ is the number of super-particles represented by the i[th] computational particle and $c_i$ is the compression factor of the i[th] particle. The $S$ term represents the shape function used to represent the computational particles in both physical and velocity space. The distribution function in Eq. 5 is used in PICOS++ to calculate the various moments used throughout the calculation as presented in section 3.4 and Appendix 1.

$$f(x, \boldsymbol{v}, t) = \frac{K}{A_0} \sum_{i=1}^{N_C} a_i c_i S(x - x_i(t)) S(\boldsymbol{v} - \boldsymbol{v}_i(t))$$

$$K = \frac{N_R}{N_{SP}} \quad N_{SP} = \sum_{i=1}^{N_C} a_i \quad c_i = \frac{B(x_i(t))}{B_0}$$

Eq. 5

Using the distribution function in Eq. 5, each computational particle has the following attributes: (1) weight $a_i$, (2) compression factor $c_i$, (3) position $x_i$ and (4) velocity vector $\boldsymbol{v}_i$. These attributes are evolved in time in the PIC simulation according to the equations of motion, magnetic field profile and boundary conditions for the particles (see section 3.3). The normalization parameters used in the code is presented in Appendix-3.

## 3.2 Guiding-center equations of motion:

In the PICOS++ framework, particles are described using the Guiding-Center (GC) approximation. The GC description assumes the magnetic field can be described in the paraxial limit (Eq. 6) and leads to the concept of magnetic moment. The use of the GC approximation removes the need to evolve the particle's gyro-phase and as result can lead to important reduction in computational time when compared to the full-orbit description. The absence of the gyro-phase variable, however, requires the use of additional approximations to implement the RF operator and the associated heating as described in section 5. In the paraxial approximation for the magnetic field, the radial component is given by Eq. 6, where $r$ is the radial coordinate. This means that the angle $\gamma$ the magnetic field makes with the $x$ axis satisfies $\cos \gamma \approx 1$. The paraxial approximation for the magnetic field is satisfied provided $\frac{r}{B_x} \frac{\partial B_x}{\partial r} \ll 1$. Here we assume that the radial transport is not significant as compared to parallel transport in Proto-MPEX and MPEX.

Eq. 6

$$B_r(r, x) = -\frac{r}{2}\frac{dB_x}{dx}$$

The GC equations of motion employed in PICOS++ are presented in Eq. 7- Eq. 9, where $x_i$ is the position along the "x" coordinate of the i$^{th}$ particle, $v_{\parallel i}$ the parallel velocity of the i$^{th}$ particle, $B_i$ and $E_{\parallel i}$ the magnetic and parallel electric field at the position of the i$^{th}$ particle respectively. The term $\mu_i$ represents the magnetic moment of the i$^{th}$ particle. Note that Eq. 9 enforces conservation of magnetic moment. In PICOS++, the equations of motion are advanced in time over a time step $\Delta t$ using a 4$^{th}$ order Runge-Kutta method. The time interval $\Delta t$ is chosen such that particles, on average, move less that a single grid cell $\Delta x$ during each time step. After positions and velocities have been advanced using Eq. 7- Eq. 9, other operations are applied such as projecting charge onto the grid, volumetric sources, Coulomb collisions and/or RF heating and electric field solution. These operations are described in the following section 4 and 5.

$$\frac{dx_i}{dt} = v_{\parallel i} \qquad\qquad Eq.\ 7$$

$$m\frac{dv_{\parallel i}}{dt} = -\mu_i \frac{dB_i}{dx} - qE_{\parallel i} \qquad\qquad Eq.\ 8$$

$$\frac{d\mu_i}{dt} = 0 \quad \text{where} \quad \mu_i = \frac{mv_{\perp i}^2}{2B_i} \qquad\qquad Eq.\ 9$$

## 3.3 Computational domain and spatial grid

Particle positions and velocity vectors are defined continuously in phase-space (not constrained to a grid); however, all electromagnetic fields and ion moments are defined on a computational grid with finite resolution. The computational domain of length L is represented with a uniform grid with $N_x$ number of grid cells. Each grid cell has a width $\Delta x$ defined as a small fraction, typically 0.1 to 0.2, of the characteristic length $d_i$ (see Appendix 3, Table A1). Hence, for the example described in Appendix 3, the cell width would be defined between 4 and 9 mm.

The computational domain can be made periodic or finite in size as in the case of a linear divertor simulator. In the case of finite-sized computational domain, the left and right boundaries are labeled $x_L$ and $x_R$; hence, for a given cell width $\Delta x$ and number of grid cells $N_x$, the spatial grid which defines the cell centers $x_p$ is defined by Eq. 10. Fields and ion moments are defined at the cell centers $x_p$. Gradients are calculated also at cell centers using central differencing. Moreover, "ghost" cells are included in the computational grid to accommodate the finite size of computational particles upon introduction of "shape" functions.

$$x_p = p\Delta x + \frac{\Delta x}{2} + x_L \quad \text{where} \quad p \in [0,1 \ldots N_x - 1] \qquad\qquad Eq.\ 10$$

In addition, computational particles which exit the computational domain by exceeding the left or right boundaries ($x_L$ or $x_R$) are re-introduced into the computational domain according to a set of rules which modify the particle attributes ($a_i, A_i, x_i, \boldsymbol{v}_i$) that define each particle. These rules are chosen to simulate a variety of boundary conditions and volumetric particle sources. This is described in more detail in reference [21].

## 3.4 Moments of the distribution function

In PICOS++, moments of the distribution function are needed to: (1) calculate the electric field and (2) apply the collision operator. These moments are calculated at cell centers $x_p$ and averaged over the cell width $\Delta x$. These moments include the ion density $n$, ion flux density $\Gamma$, parallel and perpendicular ion temperature $T_\parallel$ and $T_\perp$, etc.

Using the distribution function in Eq. 5, the ion density at an *arbitrary* location $x$ is defined by Eq. 11. Averaging over a cell width $\Delta x$ using the operation in Eq. 12 leads to the ion density $n(x_p)$ defined at the grid point $x_p$ shown in Eq. 13 where $W(x_p - x_i)$ is the so-called assignment function presented in Appendix 2 (Eq. 49) [15], [17].

$$n(x) = \int_{-\infty}^{+\infty} f d^3v = \frac{K}{A_0} \sum_{i=1}^{N_C} a_i c_i S(x - x_i) \qquad Eq.\ 11$$

$$n(x_p) = \frac{1}{\Delta x} \int_{x_p - \frac{\Delta x}{2}}^{x_p + \frac{\Delta x}{2}} n(x')\, dx' \qquad Eq.\ 12$$

$$n(x_p) = \frac{K}{A_0} \frac{1}{\Delta x} \sum_{i=1}^{N_C} a_i c_i W(x_p - x_i) \qquad Eq.\ 13$$

Having calculated all the ion moments at the cell centers $x_p$, these values can be interpolated to the particle positions $x_i$ using Eq. 14, where $F$ represents any of the ion moment or field quantities and $W$ is the assignment function.

$$F(x_i) = \sum_{p=1}^{N_X} W(x_p - x_i) F_p \qquad Eq.\ 14$$

## 3.5 Solution to the electric field

PICOS++ solves the electric field at the cell centers $x_p$ using the electrostatic approximation $E = -\nabla \phi_E$, where $\phi_E$ is the electric potential. From Faraday's law, the electrostatic approximation leads to a time-*independent* magnetic field; hence, if no plasma currents are present at $t = 0$, no plasma currents will form at future times. This simplification is used at present for its simplicity and suitability for low-beta plasma such as those encountered in linear divertor simulators. As a result, only the electric field needs to be calculated and evolved in time.

In the "hybrid" PIC approach, the electric field is solved using the generalized Ohm's law (Eq. 15) and not via the Boundary Value Problem (BVP) defined by Poisson's or Maxwell's equation as in the case of fully-kinetic PIC codes. This distinction is important because the generalized Ohm's law is an algebraic expression while Poisson/Maxwell's equations are Partial Differential Equations (PDEs). The critical difference lies in the fact that BVP defined by PDEs require boundary conditions to establish a unique solution; however, an algebraic equation such as the generalized Ohm's law (Eq. 15) *not* need boundary conditions to establish a unique solution. The electric field, as defined by the Generalized Ohm's law, is a function of the moments of the distribution function as shown in Eq. 15 where $n_e$ is the electron density and $P_e$ is the electron pressure tensor, $J_p$ the current density and $U$ the bulk plasma flow.

$$\boldsymbol{E} = \frac{\boldsymbol{J_p} \times \boldsymbol{B}}{n_e} - \boldsymbol{U} \times \boldsymbol{B} - \frac{1}{n_e} \nabla \cdot \boldsymbol{P_e} \qquad Eq.\ 15$$

In the absence of plasma currents and solving the transport along the magnetic flux (field-aligned) leads to a reduced version of the Generalized Ohms law shown in Eq. 16, where $P_{e\|}$ and $P_{e\perp}$ are the electron pressure terms parallel and perpendicular to the magnetic field respectively.

$$E_\| = -\frac{1}{en_e}\left(\frac{dP_{e\|}}{dx} - (P_{e\|} - P_{e\perp})\frac{1}{B}\frac{dB}{dx}\right) \qquad Eq.\ 16$$

At present, PICOS++ uses as an isotropic description for the electron pressure and a uniform electron temperature. These simplifications lead to Eq. 17 which is used for the electric field solution in PICOS++. Eq. 17 indicates that the electric potential is described by the Boltzmann relation. The value of the electron temperature is fixed at the start of the simulation; however, this value can be coupled to a solution to the electron energy transport equation for a more complete calculation with a time-dependent and non-uniform profile [20]. At present, the electric field in PICOS++ is calculated using Eq. 17 using central differences.

$$E_\| = -\frac{T_e}{n_e}\frac{dn_e}{dx} \qquad Eq.\ 17$$

# 4 Monte-Carlo based physics operators in PICOS++

Two Monte-Carlo based operators are used in PICOS++: (1) a Coulomb collision operator and (2) a Quasi-linear RF heating operator. These are described next.

## 4.1 Coulomb Collision operator based on Fokker-Planck equation

The Vlasov equation describes the evolution of a collection of charged particles which interact via *smooth* and *long-range* electromagnetic fields over length scales greater than the Debye length. This formulation does not include the effect of Coulomb collisions (sub-Debye length field fluctuations). Moreover, in fully ionized plasmas, small-angle cumulative Coulomb deflections is the dominant collisional process and is described via a Coulomb collision operator based on Fokker-Planck (FP) equation [49].

PICOS++ includes Coulomb collisions via a Monte-Carlo based FP collision operator based on references [11, 22, 50]. Moreover, the FP operator is applied in the *plasma* reference frame. If we let $\boldsymbol{v_i}$ and $\boldsymbol{U}$ represent the i[th] particle velocity and plasma bulk velocity vector respectively in the *laboratory* frame, the i[th] particle velocity $\boldsymbol{w_i}$ in the *plasma* frame is given by Eq. 18; hence, the particle's pitch angle $\theta_i$ is given by Eq. 19, where $\hat{\boldsymbol{b}}$ is the magnetic field's unit vector.

$$\boldsymbol{w_i} = \boldsymbol{v_i} - \boldsymbol{U} \qquad \text{Eq. 18}$$

$$\cos\theta_i = \frac{\boldsymbol{w_i} \cdot \hat{\boldsymbol{b}}}{|\boldsymbol{w_i}|} \qquad \text{Eq. 19}$$

The FP operator scatters the kinetic energy $E$ and cosine of the pitch angle $\xi = \cos\theta$ (Eq. 19) of the $i^{th}$ particle against all background species. This is performed in the *plasma* reference frame (Eq. 18). The change in kinetic energy $\Delta E_{ijk}$ and cosine of pitch angle $\Delta \xi_{ijk}$ due to the $i^{th}$ particle colliding with the $j^{th}$ background species during the $k^{th}$ time step are presented in Eq. 20 and Eq. 21 respectively.

$$\Delta E_{ijk} = -2\nu_{ij}^E \Delta t \left[ E_{ik} - \left(\frac{3}{2} + \frac{E_{ik}}{\nu_{ij}^E}\frac{d\nu_{ij}^E}{dE}\right) T_j \right] \pm 2\sqrt{T_j E_{ik} \nu_{ij}^E \Delta t} \qquad \text{Eq. 20}$$

$$\Delta \xi_{ijk} = -\xi_{ik} \nu_{ij}^D \Delta t \pm \sqrt{(1-\xi_{ik}^2)\nu_{ij}^D \Delta t} \qquad \text{Eq. 21}$$

The terms $\nu_{ij}^E$ and $\nu_{ij}^D$ represent the energy loss (Eq. 22) and deflection rates (Eq. 23) the $i^{th}$ particle colliding with the $j^{th}$ background species. $m_j$ and $T_j$ represent the mass and effective temperature of the background species $j$. The collision frequency term $\nu_{ij}^0$ is given in Eq. 24. The so-called 'Chandrasekhar' function $\psi(x)$ and the error function $\phi$ are given in Eq. 25. The term $w_{Tj}$ represents the thermal velocity of the background species $j$ and $|\boldsymbol{w_i}|$ the magnitude of the velocity vector of the $i^{th}$ particle in the *plasma* reference frame.

$$\nu_{ij}^E = \nu_{ij}^0 \left( 2\left(\frac{m_i}{m_j}\right)\frac{\psi(x)}{x} \right) \qquad \text{Eq. 22}$$

$$v_{ij}^D = v_{ij}^0 \left( \frac{\phi(x) - \psi(x)}{x^3} \right) \qquad \text{Eq. 23}$$

$$v_{ij}^0 = \frac{n_j e^4 Z_i Z_j \ln \Lambda}{2\pi m_i^2 \epsilon_0^2 w_{Tj}^3} \qquad \text{Eq. 24}$$

$$x = \frac{|\mathbf{w}_i|}{w_{Tj}} \quad \psi(x) = \frac{\phi - x(d\phi/dx)}{2x^2} \quad \phi = \frac{2}{\pi} \int_0^x \exp(-\eta^2)\, d\eta \qquad \text{Eq. 25}$$

The final kinetic energy $E_i$ and cosine of pitch angle $\xi_i$ of the $i^{th}$ particle is calculated by summing the scattering from all background species $j$, including self-collisions and collisions with the electron fluid (Eq. 26). Once the collision operator calculation is completed, the new velocity vector $\mathbf{w}_i$ is translated back to the *laboratory* frame to get $\mathbf{v}_i$ by inverting Eq. 18.

$$E_{i,k+1} = E_{ik} + \sum_j \Delta E_{ijk} \qquad \xi_{i,k+1} = \xi_{ik} + \sum_j \Delta \xi_{ijk} \qquad \text{Eq. 26}$$

It is important to note that the "background" species terms $(n_j, T_j)$ referred to in Eq. 22 to Eq. 25 are calculated from the ion moments which have been evolved self-consistently by the code (Eq. 39 to Eq. 46) and/or the electron temperature at the present time step $k$. In this manner, the FP collision operator is non-linear since it depends on the present state of the *moments* of the distribution function; however, since it uses the *moments* rather than the distribution function itself, the operator does not capture the non-linear effects due to anisotropy in velocity space. To overcome this limitation, we plan to use the fully non-linear binary-collision FP operator described in reference [51]

In PICOS++, the FP operators in Eq. 20 and Eq. 21 are sub-cycled within the simulation time interval $\Delta t$ in order to satisfy the Monte-Carlo condition. This is done by dividing $\Delta t$ into $N_*$ substeps to produce the subcycle time interval $\delta t = N_* \Delta t$ to ensure the conditions $v_{ij}^D \delta t \ll 1$ and $v_{ij}^E \delta t \ll 1$ are always satisfied. This is especially important for the low temperature plasmas to be encountered in MPEX. When these conditions are satisfied, the Monte Carlo operators are well behaved.

## 4.2 Quasilinear RF heating operator

In the GC approximation, the gyro-phase of particles is not evolved. In order to describe the resonant cyclotron interaction between charged particles and RF fields, a so-called quasilinear RF operator is implemented [11, 52, 53]. This RF operator effectively models the cyclotron resonant RF heating process as diffusion in velocity space when particles satisfy the cyclotron resonance condition shown in Eq. 27, where $n$ is the harmonic number, $\Omega_i$ the cyclotron frequency of the ith particle, $\omega_{RF}$ the RF angular frequency, $k_\parallel$ the parallel wavenumber of the RF wave and $v_{\parallel i}$ the parallel velocity of the particle in the laboratory frame.

$$n\Omega_i = \omega_{RF} - k_\parallel v_{\parallel i} \qquad \text{Eq. 27}$$

For each computational particle, an attribute called "resonance number" is calculated based on Eq. 27 in the form of $g_i = n\Omega_i + k_\| v_{\|i} - \omega_{RF}$, where $i$ represents the particle index. Whenever the attribute $g_i$ changes sign between two consecutive time steps, the particle is flagged and considered to have satisfied the cyclotron resonance condition. Particles in resonance receive a change in kinetic energy in both the perpendicular and parallel degree of freedom according to the Monte-Carlo rules given in Eq. 28 to Eq. 29 [11] The term $\Delta E_i^{RF}$ represents the mean change in kinetic energy or "RF kick" driven by the resonant interaction between the $i^{th}$ particle and the RF electric field and $R_m$ is a random number between $\pm 1$.

$$\Delta E_{\perp i} = \Delta E_i^{RF} + R_m \sqrt{2 E_{\perp i} \Delta E_i^{RF}} \qquad Eq.\ 28$$

$$\Delta E_{\| i} = \left(\frac{k_\| v_{\| i}}{n \Omega_i}\right) \Delta E_{\perp i} \qquad Eq.\ 29$$

The mean "RF-kick" term $\Delta E_i^{RF}$ in Eq. 28 is given in Eq. 30, where $E_{\pm i}$ represents the amplitude of the RF electric field at the particle position, $n$ is the harmonic number of the interaction, $k_\perp$ the perpendicular wave number of the wavefield, $r_{Li}$ the gyro radius of the particle and $\tau_a$ the RF interaction time. This last term ($\tau_a$) quantifies the time the particle spends in resonance with the RF wavefields based on the parallel velocity of the particle and the gradient of the magnetic field. Whenever $\dot{\Omega}_i$ is finite, such as away from turning points or in non-uniform magnetic fields, the RF interaction time is given by Eq. 31. Near turning points and/or near perfectly uniform magnetic fields, the interaction time is given by another expression as described in section 4.3 of reference [11].

$$\Delta E_i^{RF} = \left(\frac{e^2}{m}\right) \left(\frac{|E_{+i} J_{n-1}(k_\perp r_{Li}) + E_{-i} J_{n+1}(k_\perp r_{Li})|}{\sqrt{2}} \tau_i\right)^2 \qquad Eq.\ 30$$

$$\tau_i^2 = \frac{2\pi}{|n\dot{\Omega}_i|} \quad \text{where} \quad \dot{\Omega}_i = v_{\| i} \frac{d\Omega_i}{dx} \qquad Eq.\ 31$$

Using Eq. 28 and Eq. 29 and summing over all computational particles, we can approximate the total absorbed RF power in the plasma by the expression in Eq. 32. This expression is a very good approximation whenever a large number of particles is used and the following condition is satisfied: $E_{\perp i} \ll \Delta E_i^{RF}$. In Eq. 32, $K = N_R/N_{SP}$ represents the number of real particles $N_R$ for every super-particle, $N_{SP}$ is the total number of super-particles, $N_C$ the total number of computational particles, $a_i$ is the particle weight and $f_i^{RF}$ is a flag which is equal to 1 when the particle has satisfied the resonance condition (Eq. 27) in the current time step and is zero otherwise.

$$P_{RF} \approx \frac{K}{\Delta t} \sum_i^{N_C} a_i f_i^{RF} \left(1 + \frac{k_\| v_{\| i}}{n \Omega_i}\right) \Delta E_i^{RF} \qquad Eq.\ 32$$

At the present stage of development, the RF electric field terms ($E_\pm$) in Eq. 30 are calculated based on the time-dependent absorbed RF power specified by the user. Using Eq. 32 and the value of the desired absorbed RF power, the RF wave electric field ($E_{\pm}$) is solved for at every time step. Given this power-constrained electric field, the Monte-Carlo RF operator (Eq. 28 to Eq. 31) is applied to all resonant computational particles.

## 5 Results

In this section, we present the numerical results produced using PICOS++. To explore the various physics involved: (1) fluid plasma and (2) magnetic mirror effects, PICOS++ simulation results has also been benchmarked with the Fluid code B2.5 EIRENE [15, 54] and Proto-MPEX experimental data. Next, we introduce the numerical setup used for modeling the parallel transport in MPEX with and without the use of ICH. This is followed by an exploration of the effects of ICH power on the parallel transport dynamics, plasma density profile, flow acceleration and flux at the target.

### 5.1 Comparison of PICOS++ data with Proto-MPEX experiments

In this subsection, PICOS++ simulations are carried out with Proto-MPEX helicon only plasma parameters. The PICOS++ data is then compared against existing data from past experiments and fluid modelling on Proto-MPEX.

Benchmarking PICOS++ against fluid code B-2.5-EIRENE and Proto-MPEX Helicon only experiments

Here, we benchmark PICOS++ against fluid modelling (B2.5-EIRENE) and experimental mreasurements on Proto-MPEX under helicon-only operation. The fluid modelling and experimental details used for this exercise are presented in reference [54]. These plasmas have temperature range of 1-6 eV and thus are expected to be strongly collisional and fluid-like. With the use of a FP collision operator, PICOS++ is able to capture this process and is thus compared against a fluid model. PICOS++ does not include neutral gas dynamics which introduce ion-neutral drag, ion mometum loss and recombination processes. The main input parmeters to PICOS++ are electron density, electron temperature, magnetic field geometry and source rate which are taken from reference [54]. The boundary conditions for PICOS++ modeling are similar to those used in B2.5 EIRENE calculations [54]. Using these input parametrs and boundary consitions, PICOS++ at each iteration, solves parallel electric field and generates ion distibution functions in 1D-2V phase space. The moments of the ion distribution function are then used to calculate ion density, parallel and perpendicular temperatures, parallel bulk flow etc. The detailed simulation setup for PICOS++ has also been described in section-5.2 in context of MPEX. For this benchmarking PICOS++ uses the axial electron temperature profile obtained from B2.5 EIRENE code as shown below obtained from reference [54].

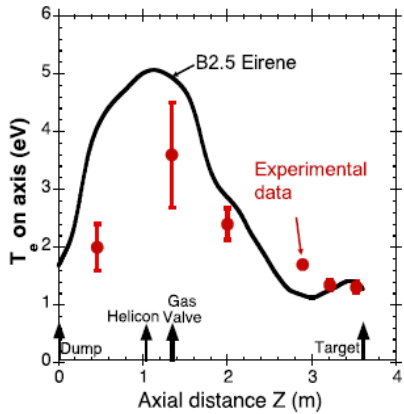

Figure 2, Axial electron temperature $T_e$ profile obtained from B2.5 EIRENE and experimental data from Proto-MPEX.

Using the electron temperature profile shown in Figure 2 as initial condition, we performed PICOS++ simulations with parameters equivalent to those used in B.2.5-EIRENE. The main plasma profiles from the PICOS++ simulations, compared with B2.5-EIRENE and Proto-MPEX experimental data are presented in Figure 3 and Figure 4.

In Figure 3, the (a) electron density $n_e$ and (b) parallel flow $U_\parallel$ profile along the axial length of Proto-MPEX are presented. Details are provided in the caption. In general, the density profiles produced by PICOS++ and B2.5-EIRENE are in good agreement, expect near the target region. Moreover, the parallel flow near the target region is overpredicted. The mismatch between the models near the boundaries could be related to neutral gas dynamics, charge-exchange and recycling at the boundaries which is not presently included in PICOS++ where the fluid modelling with B2.5 EIRENE can capture these.

In Figure 4, the (a) electron pressure $P_e$ and (b) the parallel energy flux $Q_\parallel$ along the length of the device are presented. It is evident that the axial electron pressure $P_e$ and parallel energy flux $Q_\parallel$ profiles obtained from PICOS++ (thick black curve) are in reasonable agreement with the B2.5 EIRENE and experimental data.

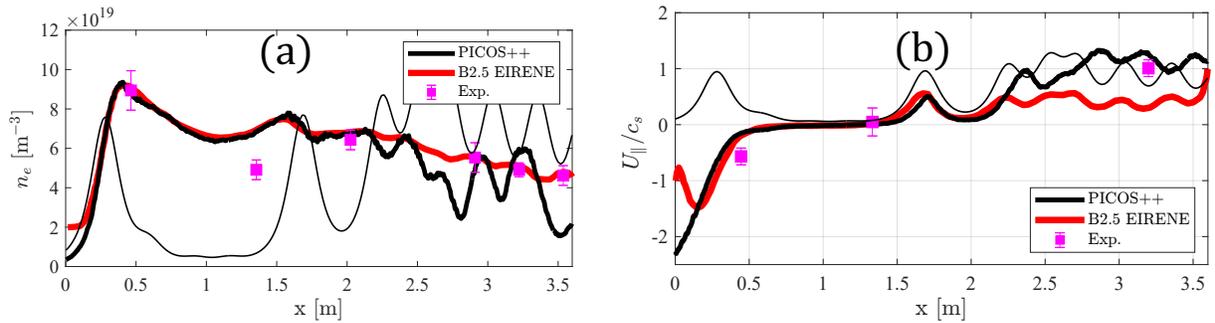

*Figure 3, (a) electron density (b) parallel flow profiles along the length of Proto-MPEX based on PICOS++ (thick black curve), B2-5-EIRENE (thick red curve) and experimental data from Proto-MPEX (magenta boxes). The thin black curve is the axial magnetic field profile superimposed on these subplots for reference.*

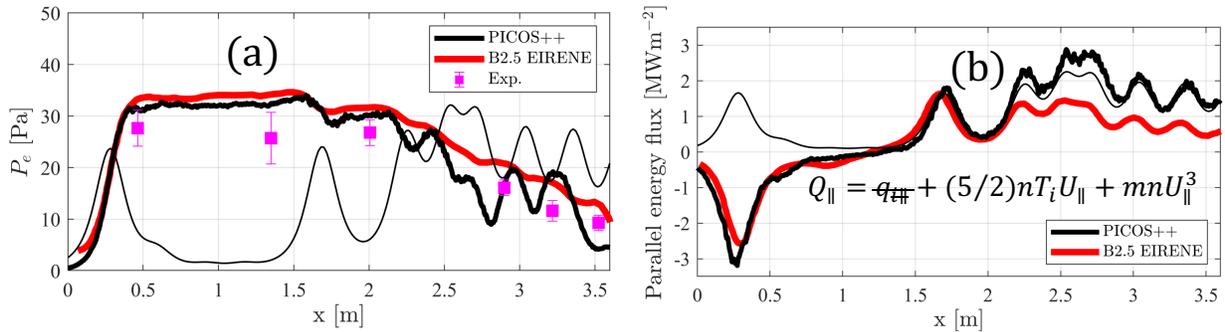

*Figure 4, (a) electron pressure and (b) parallel energy flow profiles along the length of Proto-MPEX based on PICOS++ ( thick black curve), B2-5-EIRENE (thick red curve) and experimental data from Proto-MPEX (magenta boxes). The thin black curve is the axial magnetic field profile superimposed on these subplots for reference.*

Comparison against Proto-MPEX experimental data for the helicon only case:

The magnetic field profile in Proto-MPEX is highly non-uniform because of the requirement of various cyclotron resonance based auxiliary heating schemes [11, 26]. This nonuniformity in the magnetic field leads to formation of magnetic mirrors traps at various location and thus, are expected to significantly affect the plasma parallel transport at the towards the target through the heating regions. An experimental study on Proto-MPEX helicon only conditions is performed to understand the plasma transport by systematically varying the magnetic field so as to change the mirror ratio as shown in Fig. 3, Reference [26].

A similar study with experimental setup and various input plasma parameters are described in Reference [26] has been done with PICOS++ to validate the experimental findings. The density measured at the target and at the source using PICOS++ modelling is compared against experimental data for this mirror ratio $R$ scan as shown in Figure 5. With the increase in mirror ratio, the plasma transport towards the target is reduced because of the mirror trapping of particles. This, in turn, leads to an increase in the plasma transport towards the source location. PICOS++ modelling is able to capture this effect and the simulation data are in good agreement with experimental measurements [26]. It is worthy to note here that because of the low temperature and dominant Coulomb collisions, the mirror trapping is due to the gas dynamic effect.

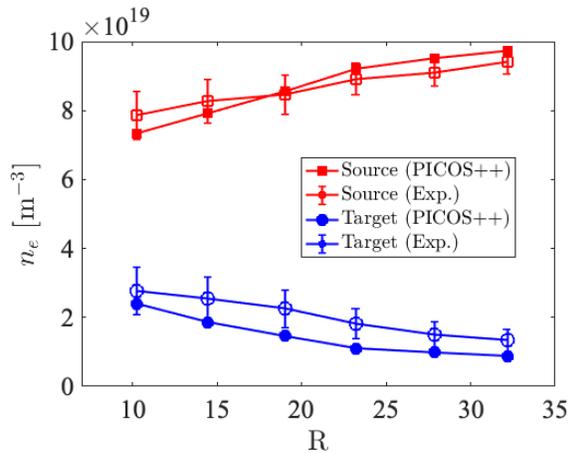

Figure 5 *The electron density measured at the target and at the source with respect to magnetic mirror ratio R in Proto-MPEX helicon only case.*

Comparison against Proto-MPEX experimental data for the helicon + ICH only case:

As also mentioned before, the Proto-MPEX is designed with the goal to perform R&D related to heating scheme on planned MPEX device. An RF based antenna is implemented in Proto-MPEX for heating the ions via ion cyclotron resonance. This is achieved by using a mode conversion scheme from the fast wave to kinetic Alfven wave. The goal of the ICH is to increase the ion flux and thereby, ion fluence reaching the target. Recent experiments using the ICH antenna have observed ion heating with temperatures up to 15 eV at the target with up to approximately 25 kW ICH power. Both the target heat flux and ion temperature increase with ICH power. More details are in reference [55]. However, during ICH operations on Proto-MPEX experiments, the electron density at the target is found to reduce with ICH power. This "density-drop" near the target is not desirable for Proto-MPEX and MPEX operations as the goal of the device is to operate with very high electron density ($n_e > 10^{21} m^{-3}$) and low electron temperature ($T_e \sim 1 - 15 eV$) similar to detached divertor condtion in a fusion reactor [56, 57]. The "density-drop" at the target as a function of ICH power obtained from simulations are compared against the experimental measurements in Figure 6. It is evident from Figure 6, that the electron density at the target $n_e$ (normalized to electron density at the target for helicon only case ($n_{e,helicon}$) obtained from simulations are in qualitative agreement with experimental observations. Both modelling and experiments on Proto-MPEX confirms that initially "density-drop" at the target increases with ICH power and shows signs of saturating beyound a certain threshold power as seen in Figure 6. The physics expalnation of this behavior is presented in detail in section-5.4.

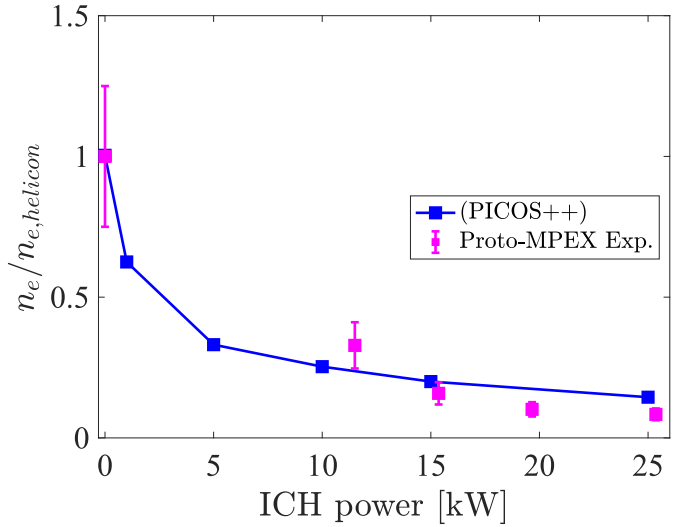

Figure 6 *The electron density $n_e$ measured at the target (normalized to electron density $n_{e,helicon}$ at the target during helicon only case) from PICOS++ modelling (blue curve) and from Proto-MPEX experiment with respect to ICH power.*

## 5.2 MPEX simulation setup

The GC based hybrid PIC code PICOS++ can be used to model any system with open magnetic geometries where the physical dimension along the magnetic field needs to be retained such as in the SOL of fusion devices, linear divertor simulators, mirror devices and plasma thrusters. In the present study, PICOS++ is applied to a linear divertor simulator-MPEX. The magnetic field profile and nominal plasma conditions are taken from the experimental data on Proto-MPEX. The magnetic field profile is presented in Figure 7. Because of the lack of neutral and atomic physics model and other assumptions, the goal here is to use PICOS++ to understand the parallel transport physics on MPEX. The goal here is not to quantitatively predict the density and temperature at the target. PICOS++ in its current formulation does not have the atomic and neutral physics to do so.

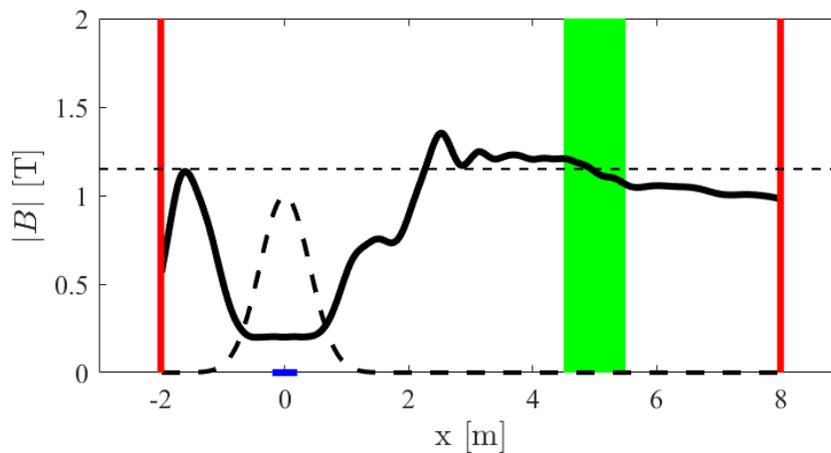

*Figure 7, The axial magnetic field profile for MPEX (thick black curve). The tick red lines on each side are the location of left and right boundaries, the shaded green region is the location the ICH power is applied. The thin dashed horizontal line is the value of the magnetic field for ICH resonance. The thick black curve centered around $x = 0$ m is the source location.*

In Figure 7 the axial magnetic field profile is represented by a thick black curve. The two red lines at $x = -2$ m and at $x = 8$ m are the representative of the left and right boundaries respectively. Particles leaving these two boundaries are re-injected back to the source location centered around $x = 0$ m with a completely new identity via a set of rules as described in Reference [21]. These reinjected particles are used as fueling source with new computational weight and with a gaussian spread of 0.3 m in physical space (see the thick dashed curve at $x = 0$ in Figure 7) and with random pitch angle in velocity space. As seen in the Figure 7, the resonance ($\omega_{RF} = \omega_{ci}$) occurs around $x = 5$ m where the magnetic field cuts the resonance line. Therefore, the ICH power in the simulation is applied between 4.0 m and 5.5 m as shown by the shaded green region. The initial conditions used in PICOS++ simulations for MPEX are shown in Table 2.

*Table 2: Simulation parameters for Hybrid PIC modeling of MPEX*

| Simulation parameters | Physical Value |
|---|---|
| Length of domain $L$ | 10 m |
| Total simulation time | 4.8ms |
| Peak magnetic field $B_m$ | 1.2T |
| Magnetic field at the source $B_0$ | 0.07T |
| Plasma radius at the source $R_0$ | 0.05m |
| Electron temperature $T_e$ | 15eV |
| Ion temperature $T_{i\parallel} = T_{i\perp}$ | 15eV |
| Ion charge number $Z_\alpha$ | 1 |
| Ion mass number $A_\alpha$ | 2AMU |
| Ion density fraction | 1 |
| Left boundary location $L_L$ | -2 m |
| Right boundary location $L_R$ | +8 m |
| Number of computational particles per cell | 2500 |
| Time step | $0.5\omega_{pi}^{-1}$ |
| Spatial grid size | 0.02m |
| Source temperature $T_{source}$ | 15eV |
| Source spread length $\sigma_{source}$ | 0.3m |
| Absorbed ICH power $P_{RF}$ | 5kW |
| Harmonic number $n$ | 1 |
| RF frequency $\omega_{RF}$ | 8.765MHz |
| Location left of the resonance $RF_{x1}$ | 4.5m |
| Location right of the resonance $RF_{x2}$ | 5.5m |
| RF on time $RF\_t\_on$ | (1.0 ms) |
| RF off time $RF\_t\_off$ | (3.8 ms) |
| Parallel wave number $k_\parallel$ | 20 m$^{-1}$ |
| Perpendicular wave number $k_\perp$ | 100 m$^{-1}$ |

## 5.3 MPEX: Helicon-only discharges

Before discussing discharges with RF heating, PICOS++ is first verified using helicon-only discharges on MPEX. The PIC simulation is evolved to steady state using a constant ion fueling at a rate of $G = 1 \times 10^{22}$ ions per second at a temperature of 15 eV that is expected on MPEX [2] and the spatial distribution shown in Figure 7. Integrating over the entire device as a function of time produces the time-evolution of the total number of ions $N_R$ in the computational domain (Eq. 33); this is shown in Fig. 3(a). In addition, the total ion flux $[s^{-1}]$ and ion flux density $[m^{-2}s^{-1}]$ arriving at the target as a function of time is shown in Fig. 3(b). The data indicate that the plasma has reached equilibrium within 2 ms. This is consistent with calculated transport timescales discussed later. Given the cross-sectional area of the plasma at the target, the *mean* ion flux density during steady-state reaching the target is $3 \times 10^{24}$ $[m^{-2}s^{-1}]$.

$$N_R(t) = \int n(x',t) A(x') \, dx' \qquad \text{Eq. 33}$$

Let us now compare these numerical results with some analytical transport calculations for mirror-trapped collisional plasmas. The total number of particles in the device $N_R$ is governed by the simple differential equation shown in Eq. 34, where $G$ is the fueling rate and $\tau_c$ is the particle confinement time which represents the mean time to empty all the particles inside the volume of plasma. In the simple case of an isotropic plasma, usually a good approximation for a low temperature helicon plasma, the particle confinement time is given by Eq. 35 [58] where $R$ is the mirror ratio, $L$ is the mirror-to-mirror length and $C_s$ is the ion sound speed. This expression is only valid for plasmas which have a fully populated loss-cone distribution. Moreover, when the source $G$ is a constant in time, the solution to Eq. 34 with an initial condition of zero particles is given in Eq. 36. In steady-state, the solution becomes $N_R(t \to \infty) = \tau_c G$.

$$\frac{\partial N_R}{\partial t} + \frac{N_R}{\tau_c} = G \qquad \text{Eq. 34}$$

$$\tau_c = \frac{RL}{2C_s} \qquad \text{Eq. 35}$$

$$N_R(t) = \tau_c G \left(1 - e^{-\frac{t}{\tau_c}}\right) \qquad \text{Eq. 36}$$

Choosing, a mirror-to-mirror length $L = 5$ and the mirror ratio of 6.7 based on Figure 7, the confinement time using Eq. 35 is approximately 0.25 ms and is consistent with the rise time of the numerical data presented in Figure 8a. Moreover, using the fueling rate of $G = 1 \times 10^{22}$ $s^{-1}$ and the calculated confinement time of 0.25 ms, we get $N(t \to \infty) = 3.15 \times 10^{18}$. In fact, the red dashed line in Figure 8a is in fact Eq. 36 and is consistent with the numerical results. This comparison demonstrates that the simulated helicon plasma behaves as expected from simple analytical transport estimates for mirror-trapped collisional plasmas.

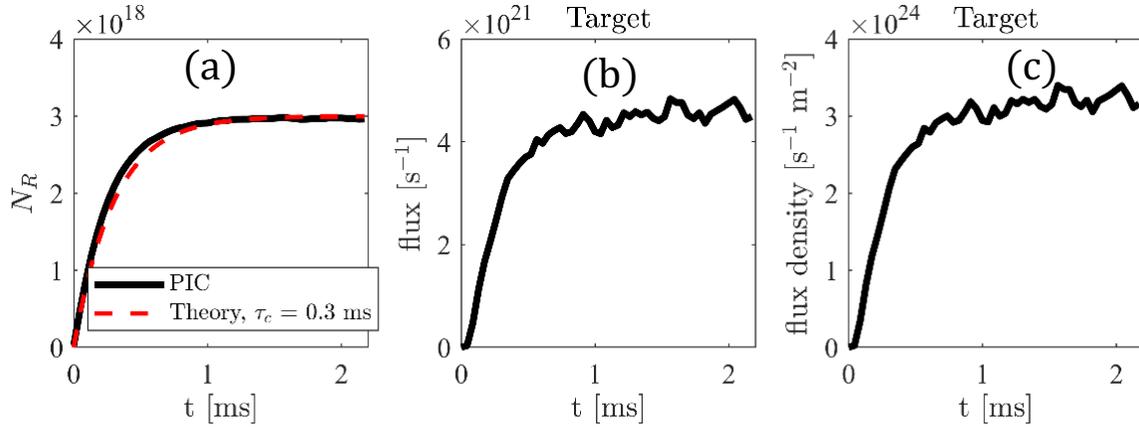

*Figure 8, (a) time-evolution of the total number of real particles $N_R$ in the simulation added to the simulation domain with; (b) Particle flux increment with time and (c) flux density with respect to time for a given constant fueling rate of $G = 1 \times 10^{22}$ ions per second.*

The steady-state plasma density $n_e$ and plasma parallel flow $U_\parallel$ (normalized to ion sound speed $C_s$) profiles calculated by PICOS++ are presented in Figure 9(a) and Figure 9(b) respectively. The most important observation is the accumulation of density in the source region between the magnetic mirrors. This is the mirror confinement effect produced in a collisional plasma. In addition, the parallel and perpendicular ion temperature profiles are shown in Figure 10. At these temperatures and plasma densities, Coulomb collision relaxation is strong enough to fully equilibrate parallel and perpendicular temperatures. Small deviations are observed near large gradients in the magnetic field. However, from these simulations, the ion distribution is isotropic.

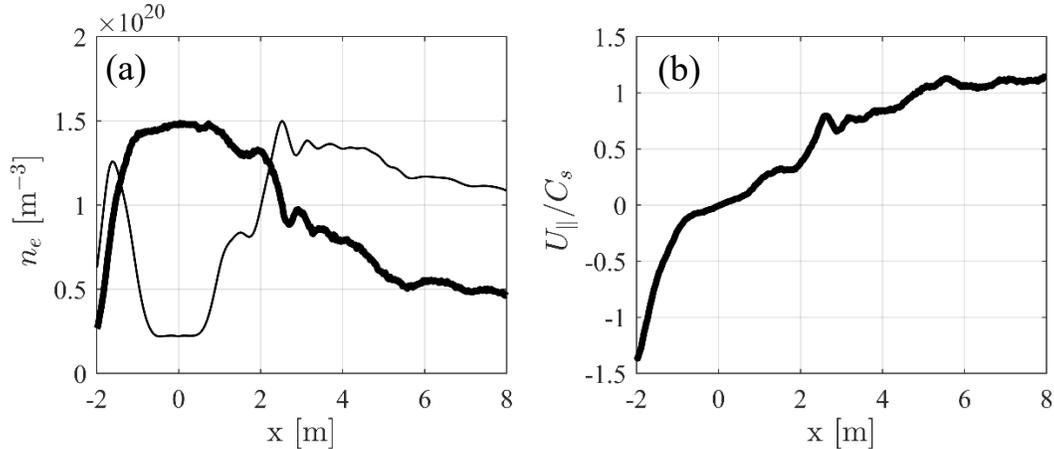

*Figure 9, (a) axial electron density $n_e$ profile for MPEX (thick black curve); (b) parallel flow $U_\parallel$ (normalized to $C_s$). The thin black curve is the axial magnetic field profile superimposed for the reference.*

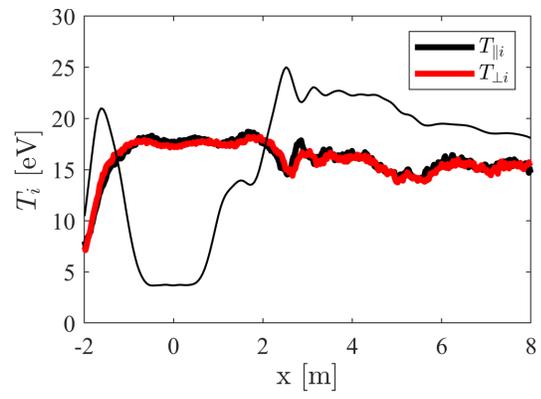

Figure 10, parallel $T_\parallel$ and perpendicular $T_\perp$ ion temperature profiles along the length of MPEX

## 5.4 MPEX: Helicon discharges with ICH

Using the "helicon-only" plasma profiles and conditions described in the previous section, ICH power is applied during the steady-state state of the helicon plasma. We now explore the effects of ion heating in MPEX.

Plasma confinement

ICH is applied to the helicon plasma once it reaches steady state (in the time range of 1 ms and 3.8 ms) as shown by the shaded green region in Figure 11. This time interval $t_{RF} = 2.8\ ms$ for ICH is sufficient for plasma to reach steady state. This can be understood from the fact the that $t_{RF} \gg \tau_c, \tau_\parallel$. Here $\tau_c = 1/v_0^{ij}$ and $\tau_\parallel = L/v_{\parallel,antenna}$ are ion-ion collisional time and ions parallel transport time respectively, $L$ being the length between antenna and the target and $v_{\parallel,antenna}$ is the parallel bulk velocity measured at the ICH antenna. For electron-ion, electron-electron, and ion-ion Coulomb collisions, the 90° cumulative scattering rate $v_0^{ij}$ of species $i$ colliding on a background species $j$ is approximately given by equation 24. For a typical ion temperature of $T_i = 473\ eV$ (maximum ion temperature observed in the simulation for 400kW ICH power ) and ion density $n_i = 4 \times 10^{19} m^{-3}$, $\tau_c \sim 0.2\ ms$ and $\tau_\parallel \sim 0.05\ ms$ which is very small compared to $t_{RF}$. The rise time in Figure 11 when ICH is applied is also consistent with these calculated timescales. Because $t_{RF}$ is approximately 14 times higher than the slowest possible time scale $\tau_c$ in the simulation, plasma reaching a steady state during ICH heating in MPEX is very well justified.

The plasma reaching steady state during ICH can also be seen from Figure 11. Figure 11 presents the time-evolution of (a) total number of particles in the entire device and (2) ion flux arriving at the Target plate with and without 100 kW of applied ICH. It is visible from the Figure 11(a) that the total number of ions increases during the heating process. In other words, application of ICH increases the confinement time of the plasma. This becomes more evident from Figure 11(b) where the ion flux at the target is reduced during ICH; hence, less particles are arriving at the target. An increase in ion flux is observed at the dump, albeit, later in the ICH pulse. This indicates that that in the present configuration, application of ICH reduces the ability of the plasma source to provide particles to the target and favors transport of particles to the dump.

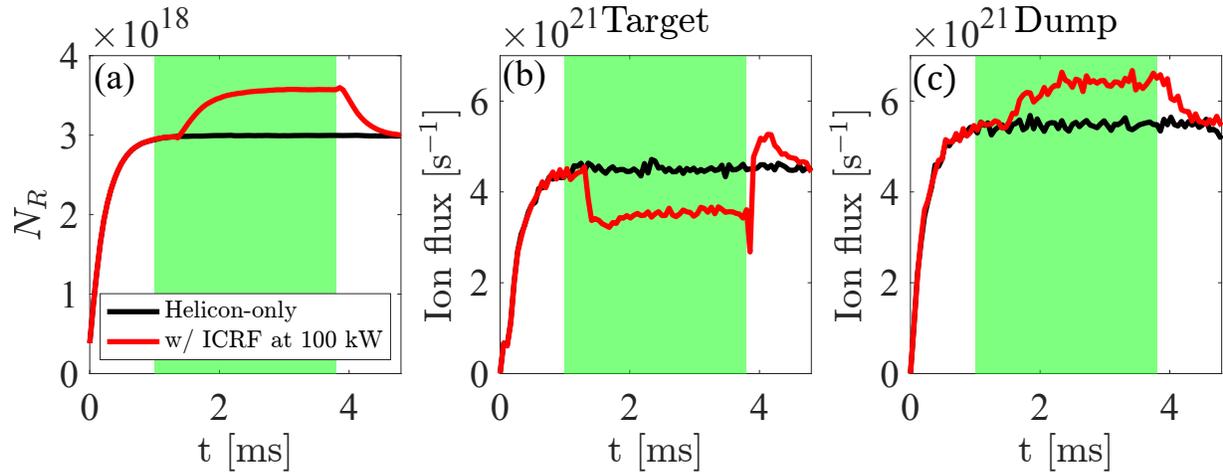

Figure 11, Time-evolution of (a) total number of particles in entire device, (2) ion flux arriving at the Target plate with and without 100 kW of applied ICH heating and (3) ion flux arriving at the dump plate with and without 100 kW of applied ICH heating.

In Figure 12, the (a) equilibrium electron density $n_e$ and (b) parallel flow $U_\parallel$ (normalized to ion sound speed $C_s$) profiles along the length of MPEX during the application of the ICH are presented. The thick black curves are the "helicon-only" profiles while the red curves the "helicon+ICH" profiles. The shaded green region represents the location where the ICH power is applied near a resonance.

The result in Figure 12(a) indicate a strong re-distribution of the plasma density, especially in the region near the resonance where the ICH power is absorbed. A very strong density gradient is formed and the density at the target is reduced. This observation is similar to the "density-drop" effect we have observed experimentally in Proto-MPEX during ICH.

Figure 12(b) indicates that the plasma flow velocity is also significantly affected during the application of ICH. In fact, the ion flow velocity slows down upstream of the ICH absorption region but it experiences a very strong acceleration downstream of the ICH region towards the target. Since no particle sources have been applied in ICH region, conservation of mass dictates that in increase in parallel flow velocity leads to a reduction in plasma density.

Moreover, application of ICH power leads to strong ion temperature anisotropy illustrated in Figure 13. In the "helicon-only" case (black thick line), because of very high collisionality, there is no temperature anisotropy observed. application and absorption of ICH, leads to sharp increase in perpendicular temperature and via collisional relaxation between degrees of freedom, the parallel temperature increases near the target region.

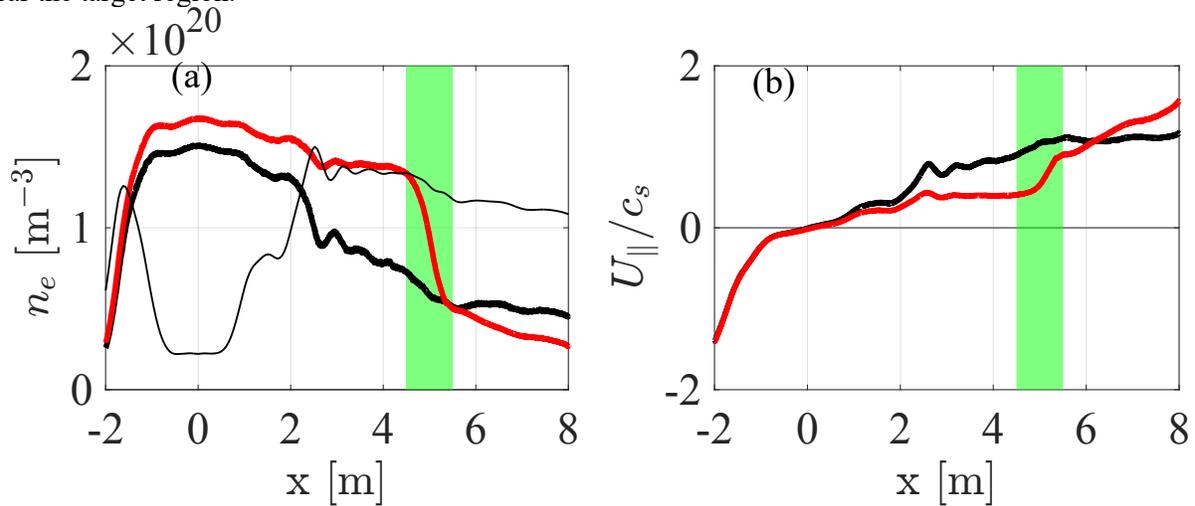

*Figure 12, (a) electron density $n_e$ (b) parallel flow $U_\parallel$ (normalized to ion sound speed $C_s$) profiles along the length of MPEX at 3.8 ms. The thick black curves represent the "helicon-only" profiles, while the red curves the "helicon+ICH" profiles. The shaded green region represents the location where the ICH power is applied near the resonance. The thin black curve is the axial magnetic field profile superimposed on these subplots for reference.*

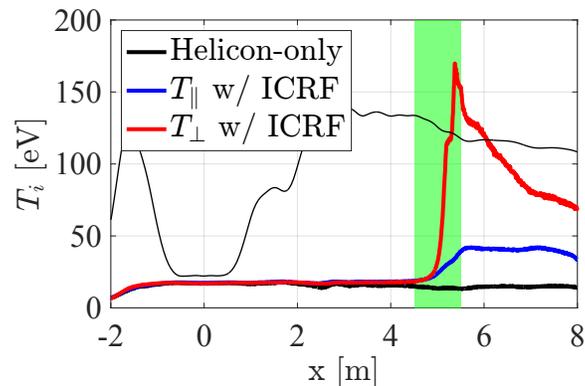

*Figure 13, Parallel $T_\parallel$ and perpendicular $T_\perp$ temperature profiles along the length of MPEX at 3.8ms. The thick black curve is the "helicon-only" temperature profile. The thin black curve is the axial magnetic field profile superimposed on these subplots for reference.*

Ion distribution function

Figure 14 shows the distribution functions at the source, ICH and target regions during steady-state. This figure confirms that the ion distribution function at the source is essentially isotropic and Maxwellian. Moreover, the distribution function at the target has a significant drift velocity but is "stretched" in the perpendicular degree of freedom indicating a two-temperature distribution as evidenced in Figure 14. However, in the ICH region the distribution function is significantly anisotropic.

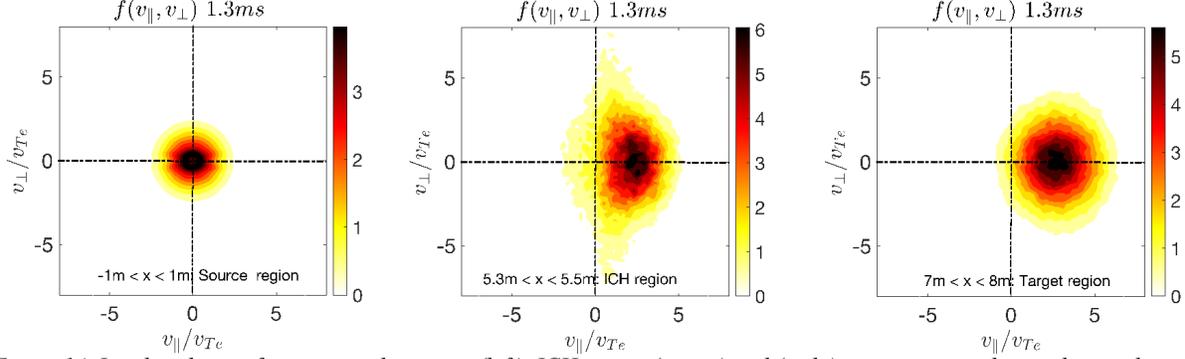

*Figure 14, Ion distribution functions at the source (left), ICH region (center) and (right) target regions during the steady-state period (at 3.8ms) when the ICH power is applied.*

Force balance

The steady-state momentum transport equation along the magnetic flux [21] is given in Eq. 37. This expression includes the effect of magnetic compression arising due to a non-uniform magnetic field. In Eq. 37, first term on left hand side ($f_K = B\frac{\partial}{\partial x}\left(\frac{MnU_\parallel^2}{B}\right)$) is the plasma acceleration force, first term ($f_\parallel = -\frac{\partial P_\parallel}{\partial x}$) on right hand side (RHS) is the parallel pressure gradient and second term ($f_B = -\frac{P_\perp - P_\parallel}{B}\frac{\partial B}{\partial x}$) on the RHS is the mirror force arising mainly due to pressure anisotropy and modulated by the parallel magnetic field profile. The individual contribution from each term $f_K$, $f_\parallel$ and $f_B$ are analyzed using the PICOS++ simulation data as shown below in Figure 15. The green shaded region is the location of ICH resonance.

A very important observation is the strength of the magnetic force term ($f_B$) is very weak relative to the other two forces ($f_K$ and $f_\parallel$). This means that despite the strong ion temperature anisotropy (Figure 14), the mirror force in the present situation is not important. This is mainly due to the uniformity of the magnetic field near the ICH region.

The numerical results demonstrate that the force balance is primarily between the parallel pressure and the plasma acceleration forces while the mirror force, even when significant temperature anisotropy exists, is only of secondary importance. The ICH heating directly accelerates the bulk plasma towards the target. This acceleration is balanced by reaction force which pushes the bulk plasma away from the target via the pressure gradient force. These forces help to explain the "density drop" observed at the Proto-MPEX and MPEX target.

$$B\frac{\partial}{\partial x}\left(\frac{MnU_\parallel^2}{B}\right) = -\frac{\partial P_\parallel}{\partial x} - \frac{P_\perp - P_\parallel}{B}\frac{\partial B}{\partial x} \qquad Eq.\ 37$$

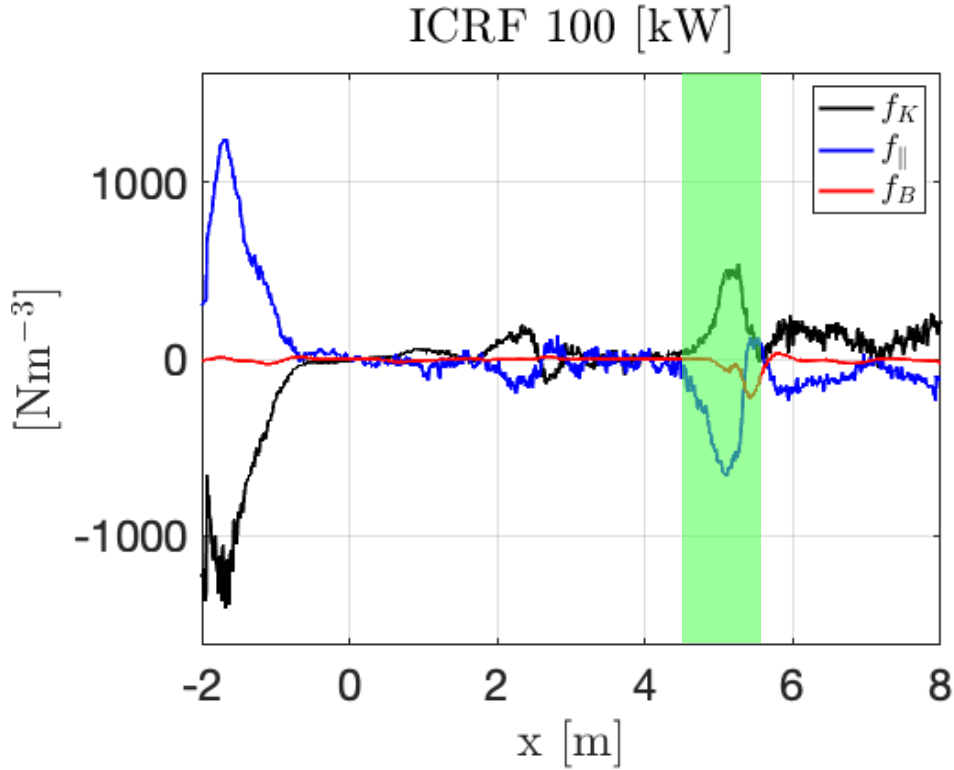

*Figure 15, Axial variation of parallel kinetic force $f_k$ (black curve), parallel pressure gradient force (blue curve) and magnetic mirror force along the length the MPEX device at 3.8 ms.*

## 5.5 Scaling with ICH power

The results presented in the previous section indicate that the plasma flux towards the target is reduced upon application of ICH power. Moreover, it indicated that the plasma flux was increased towards the dump. The main question now is how these two effects scale with increasing ICH power. In this section, we investigate precisely this question by systematically increasing the ICH power from 0 to 400 kW on MPEX while maintain all other conditions fixed. The simulation setup is identical to that presented in the previous section.

Plasma density and parallel flow
Firstly, we present the plasma density and flow velocity upstream and downstream of the ICH region as these are variables that are most strongly affected during ICH (see Figure 12). Moreover, we also present the parallel and perpendicular ion temperatures at the target in order to quantify the level of anisotropy as a function of ICH power. The results are presented in Figure 13.

During application of ICH power in MPEX, the plasma density (Figure 16a) upstream of the ICH resonance increases but saturates to a maximum level at about 100 kW. The density drop at the target follows a similar trend. In other words, increasing the ICH power beyond 100 kW in these cases does not leads to further reduction in plasma density at the target. On the other hand, the plasma flow velocity is observed to continuously accelerate as we increase the ICH power (Figure 16b); however, the plasma flow reduction upstream of the ICH region reaches saturation at about 100 kW.

The parallel and perpendicular ion temperatures at the target as a function of ICH power are presented in Figure 16c. The perpendicular component increases linearly with ICH power while the parallel component saturates for ICH powers greater than 200 kW. Since the parallel ion temperature is driven to large extent by the collisional relaxation of the perpendicular degree of freedom, as the ICH power is increased, the higher temperatures means that relaxation takes longer to occur and the increases plasma flow towards the target means that there is less time available for the ions to isotropize before they reach the target plate.

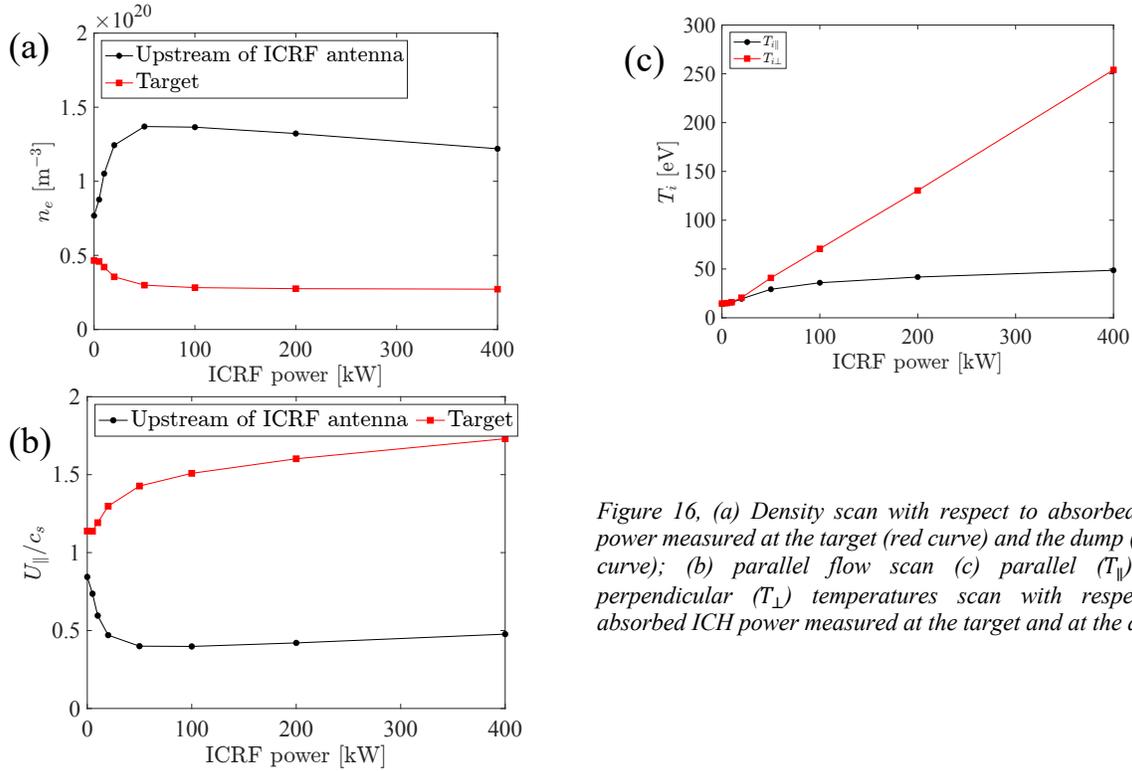

Figure 16, (a) Density scan with respect to absorbed ICH power measured at the target (red curve) and the dump (black curve); (b) parallel flow scan (c) parallel ($T_\parallel$) and perpendicular ($T_\perp$) temperatures scan with respect to absorbed ICH power measured at the target and at the dump.

Ion flux towards the target

Given that the plasma density and flow velocity at the target scale inversely to each other as a function of ICH power, a critical question is what happens to the ion flux at the target as a function of ICH power?

In Figure 17, the (a) total particle and (b) energy fluxes to the target and dump plates are presented. Figure 17(a) indicates that while the total ion flux towards the target drops, as observed in Figure 17b, it saturates at about 100 kW. The same is true for the total ion flux at the dump plate. What this means is that increasing the ICH power beyond 100 kW does not lead to a further decrease in particle transport towards the target. This observation is very important as it suggests a potential pathway to circumvent density drops that may occur in MPEX. More details on this are presented in the discussion.

Finally, the power flux towards the target and dump as a function of ICH power are presented in Figure 17(b). The results clearly demonstrate that most the ICH power absorbed is effectively coupled to the target. The other fraction of power is lost to electron-ion collisions. This collisional power should lead to an increase in electron temperature; however, this process is currently not available in PICOS++. We note that almost no ICH power is transported to the dump plate.

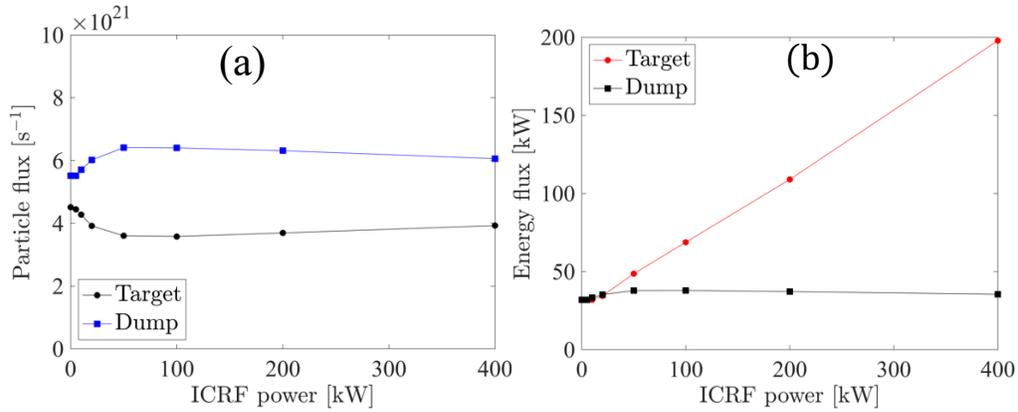

*Figure 17, (a) Particle and (b) power flux with respect to absorbed ICH power measured at the target (red curve) and the dump (black curve).*

Analysis on plasma transport time scales

During ICH power scan from 0 to 400 kW, the electron $n_e$, particle flux, parallel flow $U_\parallel$, parallel ion temperature $T_{i\parallel}$ and parallel energy flux at the target (Figure 16-17) are observed to saturate beyond 100 kW. To further understand this behavior, we inspected the variation of ion-ion collisional time $\tau_c$ and ions parallel transport times scales with respect to ICH power as shown in Figure 18. Careful analysis of Figure 18 suggests that at ~100 $kW$, the collisional and parallel transport timescales are similar. As expected at lower ICH power ($< 100\ kW$), collisions dominate over the parallel transport. However, at higher ICH power (>100 kW), the parallel transport starts to dominate over collisional transport. During the ICH heating of ions, the temperature is increased in perpendicular degree of freedom and rises in parallel temperature is mediated via collisional relaxation process. Since collisions become less significant beyond 100kW, the mechanism to covert perpendicular heating into heating in parallel degree of freedom via collisions also becomes less significant. This leads to the saturation of plasma profiles beyond 100 kW measured at the target even when the ion perpendicular temperature continues to increase.

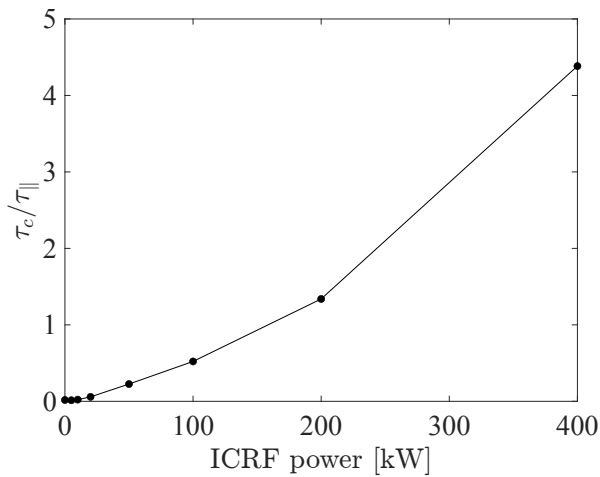

*Figure 18, The ratio between ion-ion collisional time $\tau_c$ and ions parallel transport time $\tau_\parallel$ with respect to ICH power.*

# 6 Discussion

In this work, a guiding center based 1D-2V Hybrid PIC code-PICOS++ is developed to model plasma parallel transport with (1) Coulomb collisions, (2) Quasilinear RF heating operator and (3) Volumetric particle sources with rate constaint fueling.

## 6.1 Why Hybrid PIC approach?

Kinetic-ion, quasi-neutral, fluid-electron hybrid codes occupy a unique region of simulation capability. On one hand, they model ions kinetically: the ability of the ion distributions to be non-Maxwellian allows the codes to intrinsically capture many effects that are difficult, if not impossible, to capture in a fluid-ion formulation. On the other hand, such hybrid codes are not restricted to resolving electron spatial scales such as the Debye length, the electron skin depth or the electron gyro-radius, and their associated time scales. Thus, it is possible to use these hybrid codes to model much larger spatial/temporal scales and use higher dimensionality than is practical using a fully kinetic method. However, we note that when not including kinetic electrons, we cannot model inertial electron dynamics or other electron kinetic effects. An additional advantage of hybrid over fully kinetic methods is that they have a well-defined electron temperature, and thus it is easier to interface the system with physics modules such as equation-of-state (EOS), ionization, and radiation transport.

The main purpose of developing this Hybrid PIC code- PICOS++ is to study the plasma parallel transport with ion cyclotron RF heating in the steady state linear divertor simulator-MPEX. The Hybrid PIC formalism is able to capture the evolution of distribution function primarily determined by the the interplay between RF-driven velocity space diffusion, collisional relaxation and magnetic mirror in MPEX. To test the various Monte-Carlo physics operators: Coulomb collision (Fokker-Planck equation) and quasilinear RF heating operatrors, the code is benchamrked with the existing Proto-MPEX experimental data for helicon only and helicon+ICH scenarios. PICOS++ is also benchmarked with results from the fluid code-B2.5-EIRENE. During this benchmarking exercise, the PICOS++ modeling is found consistent with the experiments and fluid simulations using B2.5-EIRENE.

## 6.2 "Density-drop" at the target in MPEX:

During the modelling with ICH heating in MPEX, density at the target *drops* significantly as compared to upstream density and constistent with past experiments observations on Proto-MPEX. This "density-drop" near the target is not desirable for MPEX operations as the goal of the device is to operate with very high electron density ($n_e > 10^{21} m^{-3}$) and low electron temperature ($T_e \sim 1 - 15 eV$) similar to detached divertor condtion in a fusion reactor [56, 57].

The most important observation from the modelling is that while both the density and flow velocity are strongly affected during ICH, the total particle flux towards the target is only weakly affected by the ICH power. In fact. the total ion flux towards the target is observed to drop by 20% and saturates at about 100 kW (Figure 17a). It is quite possible that the neutral gas recycling at the target along with the electron heating can be used to recover the "lost" density at the target. PICOS++ currently does not have a model for neutral gas recycling and charge exchange. However, in future, a fluid neutral model can be added to PICOS++ to explore these effects on "density-drop" near the target in MPEX under the effects of neutral gas recycling and charge exchange.

## 6.3 ECRH modeling for MPEX

In the present Hybrid-PIC approach, fluid electron and kinetic ions, the electron distribution function is assumed to be an isotropic Maxwellian at a temperature $T_e$

However, for ECR heated plasmas, electron distribution is no longer remains isotropic nor Maxwellian. In this case, kinetic treatment of electrons becomes essential and cannot be modeled by present version of the Hybrid PIC code-PICOS++. Fully kinetic simulations are capable of simulating ECR plasmas, however, at huge computational cost. One way to model plasma during ECH is to use a Hybrid PIC approach with *"kinetic electrons and fluid ions"* described in Reference [20, 59, 60]. In this formulation, parallel electric field $E_\parallel$ is calculated using the electron parallel momentum conservation equation shown in Eq. 38.

$$E_\parallel = -\frac{1}{en_e}\left(\frac{d(n_i T_{e\parallel})}{dx} - (n_i T_{e\parallel} - P_{e\perp})\frac{1}{B}\frac{dB}{dx}\right) \qquad Eq.\ 38$$

In the above equation, electron inertia is ignored and $P_{e\parallel} = n_i T_e$. $P_{e\perp}$ and $P_{e\parallel}$ are the electron kinetic pressure in perpendicular and parallel direction. The sheath width is set to zero and the sheath potential is calculated self-consistently by requirement of electron flow at the wall to be consistent with quasi-neutrality condition. Using this Hybrid-PIC (kinetic electrons and fluid ions) method, the quasi-linear RF heating operator to model ECR heating in MPEX can be implemented along with the Coulomb collision operator based on FP equation very similar to what described in section 4 of this paper. The details of this development in PICOS++ will be the part of next publication.

## 6.4 Particle rezoning method:

In general PICOS++ can be applied to any open magnetic field system for instance Wisconsin HTS axisymmetric mirror (WHAM) device [21, 61]. In presence of very strong magnetic mirror ratio of 34 in WHAM, the particle leaving out to the exhaust from mirror regions are very small in number. This leads to poor statistics and considerable noise in the plasma profiles. This can be handelled by either using adaptive mesh grids with finer spacing in the regions of interset or by rezoning/increasing the number of particles in regions of high accuracy and reducing the number of particles where lower accuracy can be tolerated as described in reference [62, 63]. Finer grid spacing leads to a better description of the electromagnetic fields, but particle rezoning is needed to gain a better description of the plasma dynamics and a reduction of noise. The particle rezoning is also very useful in increasing selectively the accuracy in specific velocity range. The implementation of adaptive mesh grid and particle rezoning in PICOS++ will be the subject of another paper.

## 7 Acknowledgements

This research used resources of the Fusion Energy Division, FFESD at the Oak Ridge National Laboratory, which is supported by the Office of Science of the U.S. Department of Energy under Contract No. DE-AC05-00OR22725. This research also used resources of the National Energy Research Scientific Computing Center (NERSC); a U.S. Department of Energy Office of Science User Facility located at Lawrence Berkeley National Laboratory.

# Appendix 1: Ion moments

In PICOS++, the ion moments are calculated at all cell centers $x_p$ and averaged over a cell width $\Delta x$. The expressions used to perform these calculations are presented in Eq. 39 to Eq. 42. $\Gamma_\parallel$ represents the ion parallel flux density, $P_{11}$ and $P_{22}$ the (1,1) and (2,2) terms of the ion stress tensor (see 3.17 in [51]) respectively, and $m$ is the ion mass.

$$n(x_p) = \frac{K}{A_0} \frac{1}{\Delta x} \sum_{i=1}^{N_C} a_i c_i\, W(x_p - x_i) \qquad Eq.\ 39$$

$$\Gamma_\parallel(x_p) = \frac{K}{A_0} \frac{1}{\Delta x} \sum_{i=1}^{N_C} a_i c_i\, W(x_p - x_i) v_{\parallel i} \qquad Eq.\ 40$$

$$\frac{P_{11}(x_p)}{m} = \frac{K}{A_0} \frac{1}{\Delta x} \sum_{i=1}^{N_C} a_i c_i\, W(x_p - x_i) v_{\parallel i}^2 \qquad Eq.\ 41$$

$$\frac{P_{22}(x_p)}{m} = \frac{K}{A_0} \frac{1}{\Delta x} \sum_{i=1}^{N_C} a_i c_i\, W(x_p - x_i) v_{\perp i}^2 \qquad Eq.\ 42$$

Using the above ion moments, the ion parallel drift velocity, pressures (parallel and perpendicular) and temperatures are calculated using Eq. 43 to Eq. 45.

$$u_\parallel(x_p) = \frac{\Gamma_\parallel}{n} \qquad Eq.\ 43$$

$$p_\parallel(x_p) = P_{11} - mnu_\parallel^2 \quad \text{and} \quad p_\perp(x_p) = P_{22} \qquad Eq.\ 44$$

$$T_\parallel(x_p) = \frac{p_\parallel}{n} \quad \text{and} \quad T_\perp(x_p) = \frac{p_\perp}{n} \qquad Eq.\ 45$$

Summing over all ion species, the electron density and plasma bulk flow are given by Eq. 46.

$$n_e = \sum_\alpha Z_\alpha n_\alpha \quad \text{and} \quad U_\parallel = \frac{1}{n_e} \sum_\alpha Z_\alpha n_\alpha u_{\alpha\parallel} \qquad Eq.\ 46$$

# Appendix 2: Shape and assignment function

PICOS++ uses the Triangular-Shaped Cloud (TSC) shape function given in Eq. 47 [15]. Given the position $x_i$ of a computational particle, its charge is spread along the grid according to the shape function (Eq. 47). Moreover, the fraction of the total charge assigned to an arbitrary cell center $x_p$ with a width $\Delta x$ is calculated using the integral in Eq. 48. The assignment function $W(x_p - x_i)$ associated with Eq. 47 is given in Eq. 49.

$$S(x_p - x_i) = \frac{1}{\Delta x} \begin{cases} 1 - \frac{|x_p - x_i|}{\Delta x} & \text{when } \frac{|x_p - x_i|}{\Delta x} \leq 1 \\ 0 & \text{when } \frac{|x_p - x_i|}{\Delta x} > 1 \end{cases} \qquad Eq.\ 47$$

$$W(x_p - x_i) = \int_{x_p - \Delta x/2}^{x_p + \Delta x/2} S(x' - x_i)\, dx' \qquad Eq.\ 48$$

$$W(x_p - x_i) = \frac{1}{\Delta x} \begin{cases} \frac{3}{4} - \left(\frac{|x_p - x_i|}{\Delta x}\right)^2 & ,\ 0 < |x_p - x_i| < \frac{\Delta x}{2} \\ \frac{1}{8}\left(3 - 2\frac{|x_p - x_i|}{\Delta x}\right)^2 & ,\ \frac{\Delta x}{2} < |x_p - x_i| < \frac{3\Delta x}{2} \\ 0 & \text{otherwise} \end{cases} \qquad Eq.\ 49$$

# Appendix 3: Normalization of variables

All input and output files used in PICOS++ are given in physical SI units. However, in the initialization stage of the computation, all physical quantities are normalized by a set of characteristic scales to reduce round-off errors. The characteristic scales selected for PICOS++ are the following: (1) the speed of light $c$ (characteristic velocity), (2) the inverse ion plasma frequency $\omega_{pi}^{-1}$ (characteristic time), (3) average ion charge $\bar{q}_i$ (characteristic charge) and (4) average ion mass $\bar{m}_i$ (characteristic mass). The characteristic length $d_i$ is calculated as $d_i = c\omega_{pi}^{-1}$, which is the definition of the ion skin depth.

Since the plasma modelled by PICOS++ is in general time-dependent and multi-species, the inverse ion plasma frequency $\omega_{pi}^{-1}$ is calculated using: (1) the average ion mass $\bar{m}_i$, (2) average ion charge $\bar{q}_i$ and (3) a density value characteristic of the problem under investigation. For example, in a typical MPEX deuterium plasma, the characteristic density could be chosen to be $5 \times 10^{19}$ m$^{-3}$. This leads to a characteristic length of approximately 4.5 cm. Using the main characteristic scales: (1) time, (2) velocity, (3) mass and (4) charge, the characteristic electric, magnetic fields and temperature are calculated as shown in Table-A1.

*Table A1, Characteristic plasma parameters used to normalize the simulation variables.*

| Velocity | $c$ | Light speed in vacuum |
|---|---|---|
| Time | $\omega_{pi}^{-1}$ | Inverse of ion plasma frequency |
| Mass | $\bar{m}_i$ | Average ion mass |
| Charge | $\bar{q}_i$ | Average ion charge |
| Number density | $\bar{n}_e$ | Characteristic electron density |
| Length | $d_i = c\omega_{pi}^{-1}$ | Ion skin depth |
| Electric field | $(\bar{m}_i/\bar{q}_i)c\omega_{pi}$ | |
| Magnetic field | $(\bar{m}_i/\bar{q}_i)\omega_{pi}$ | |
| Temperature | $\underline{m}_i c^2/k_B$ | $k_B$ is the Boltzmann constant |